\documentclass{PoS}
\input epsf 
\newcommand{\beq}{\begin{equation}}
\newcommand{\eeq}{\end{equation}}
\newcommand{\bea}{\begin{eqnarray}}
\newcommand{\eea}{\end{eqnarray}}

\title{Critical couplings and string tensions from two-lattice matching of RG decimations}

\ShortTitle{Critical couplings from two-lattice matching of RG decimations}

\author{X. Cheng\\
        Department of Physics and Astronomy, University of California, Los Angeles\\
        Los Angeles, CA 90095, USA\\
        E-mail: \email{darktree@physics.ucla.edu}}
\author{\speaker{E. T. Tomboulis}\\
       Department of Physics and Astronomy, University of California, Los Angeles\\
       Los Angeles, CA 90095, USA\\
       E-mail: \email{tomboulis@physics.ucla.edu}}

\abstract{Critical couplings and string tensions in $SU(2)$ and $SU(3)$ lattice gauge theory are calculated by two-lattice matching of 
RG block transformations. The transformations are of the potential moving type generating 
plaquette actions with large number of group characters and exhibit rapid approach to a unique renormalized trajectory. 
Fixing the critical coupling $\beta(N_\tau)$ at one value of temporal lattice extent $N_\tau$ 
by MC simulation, the critical couplings for other values of $N_\tau$ are then obtained by lattice matching of these decimations. $\beta(N_\tau)$ values are thus obtained for a range of $N_\tau$ 
and found to be in agreement with MC simulation results to within a few percent in all cases. A similar 
procedure allows the calculation of string tensions for a range of $\beta$ values 
with similarly good agreement with MC data. 
}

\FullConference{The 30 International Symposium on Lattice Field Theory - Lattice 2012,\\
		June 24-29, 2012\\
		Cairns, Australia}

\begin{document}

\section{Introduction}
The renormalization group (RG) based technique of  ``lattice 
matching" via block transformations relates physical quantities on different lattices. It thus provides a method for computation of a physical quantity at different lattice spacings (couplings). 
To apply it one needs to implement RG block transformations on the approach to the 
Wilsonian Renormalized Trajectory (RT). This can be done in various ways. One way is numerical 
implementation of the RG blocking by Monte Carlo RG (MCRG) techniques. This is the method that has mostly been used in the literature. Another approach is to implement blocking by explicit RG recursion relations that can, to varying degree, be explicitly carried out by numerical-analytical means. This is the method followed here. Specifically, we employ explicit RG recursion relations of the ``potential moving" type. These block transformations (decimations) are of course approximate
but can, in principle, be systematically improved.  They turn out to be surprisingly effective. 
In the following we apply lattice matching of these decimations to obtain critical couplings and string tensions for the $SU(2)$ and $SU(3)$ pure gauge theories. A more 
detailed account has appeared in \cite{CT}.

\section{RG blocking recursions and lattice matching} 
{\it RG blocking recursions}\  
We start by assuming a general plaquette action $A_p(U_p,n)$ on lattice of spacing $b^na$ given in terms of the character expansion of its exponential:  
\beq 
\exp \left(-A_p(U_p, n)\right) 
   = \sum_j\;d_j\, F_j(n)\,\chi_j(U_p) \; . \label{exp} 
\eeq
The sum is over all inequivalent irreducible representations labeled by $j$, with 
characters $\chi_j$ of dimension $d_j$.  
The action itself is, of course, completely specified by the set of $F_j(n)$ coefficients, and vice versa,  and of the general form: 
\beq
A_p(U_p,n) = \sum_j \;{1\over  d_j} \beta_j(n) \, {1\over 2 l_j}[ \chi_j(U_p) + \chi_j(U_p^{-1})]     \label{actexp}
\eeq 
with $l_j=1$ for self-conjugate and $l_j=2$ for non-self-conjugate representations. (For   
$SU(2)$, in particular, $l_j=1$ for all $j$.)

It is useful to define an effective coupling $g^{(n)}$  characterizing a given action of the form (\ref{actexp}):
\beq
\beta^{(n)} = {2N\over g^{(n)\,2}} \equiv  \left. 2N {d^2 A_p(e^{i\theta \hat{m}\cdot t}, n) \over d \theta^2}\right|_{\theta=0} \,. \label{effcoupl}
\eeq 
Here $\{t\}$ are the $SU(N)$ generators and  $\hat{m}$ a unit vector. 
((\ref{effcoupl}) is of course independent of the direction $\hat{m}$).  
In the perturbative regime this reduces to the usual definition of gauge coupling. 
In the non-perturbative regime any definition of  a `coupling' is of course some scheme-dependent choice. We adopt (\ref{effcoupl}) to track the RG recursion flows; it 
provides an efficient parametrization of the renormalized trajectory below. 

The lattice block step $b^n a \to b^{n+1}a$ may now be formulated as a prescription for 
the character expansion coefficients $F_j(n+1)$ in terms of the $ F_j(n)$'s: 
\beq 
F_j(n+1) = \left[ \int\, d U\;
\left[ \sum_k\;d_k\, F_k(n)\, \chi_k(U) \right]^{\zeta^{(d-2)}}
     \,{1\over d_j}\,\chi_j^*(U)  \right]^{r^2}  \;.   \label{recur} 
 \eeq
To complete the prescription we must specify the renormalization parameters $\zeta$, $r$. We take 
\bea
\zeta &  = & b\left[1 - c \, g^{(n)\,2} \right]          \label{decpar1} \\
r & = & b \left[ 1 - c \, g^{(n)\,2} \right]  \,      \label{decpar2}
\eea
with $c$ an adjustable decimation parameter to be tuned for optimization as explained below. 
It is convenient to work with normalized coefficients 
$f_j=F_j/F_0$ by factoring out the trivial representation coefficient in (\ref{exp}). 
%; the correspondingly normalized action differs by a trivial 
%shift of the constant (trivial character) part in (\ref{actexp}). 
Effective couplings (\ref{effcoupl})
are also conveniently computed directly in terms of the $\{f_j\}$.

{\it RG flows and lattice matching}\  
Consider now a general lattice system described by an action $A(K)$  with set of couplings $K = \{K_i\}$. Successive RG blockings  by a scale factor $b$ generate a 
flow in action space: $K \to K^{(1)} \to K^{(2)} \to \cdots \to K^{(n)} \to \cdots $, 
where  $K^{(n)}=\{K^{(n)}_i\}$ denotes the couplings after $n$ blocking steps. 
\begin{figure}[ht]
\begin{center}
\includegraphics[width=0.4\textwidth]{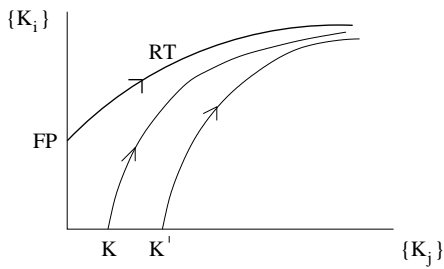}
\end{center}
\caption{Flows towards the RT from two different starting couplings $K$ and $K^\prime$.} 
\end{figure} 
If flows from $K$ and $K^\prime$ reach the same point on RT after 
$n$ and $n^\prime$ steps, then the corresponding lattice correlation lengths $\xi$, $\xi^\prime$ 
and spacings $a$, $a^\prime$ are related as  
\beq 
\xi^\prime = b^{-(n- n^\prime)} \xi \qquad 
a^\prime = b^{(n-n^\prime)} a \, . \label{latspace1}
\eeq

To identify such pairs of couplings we need ascertain that after $n$ and $n^{\, \prime}$ RG steps, respectively,  the same point is reached on the RT. 
This can be done in either of two ways: 
(i) show that the corresponding actions coincide: $A(K^{(n)}) = A(K^{\,\prime (n^{\,\prime})})$.  
This requires that one obtain the blocked action at each step; or 
(ii) show that the expectations of every operator, measured 
after performing the corresponding number of blocking steps from the initial two actions, agree. 
Either way, identifying such pairs $(K,\, n)$, and $(K^\prime, \, n^\prime)$ 
is referred to as two-lattice matching \cite{H1}.  

If blockings  are performed numerically by MCRG, the second method appears  easier to use. Obtaining the blocked action can be difficult, whereas it is possible, at least in principle, to generate a
Boltzmann-weighted configuration ensemble for the blocked action by instead blocking the configurations of an ensemble generated from the original action. These can then be used to measure observables \cite{H2}. 
In practice, of course, due to lattice size limitations, only a rather small number of block steps is possible by MCRG, so getting close enough to the RT is not guaranteed. 
As a general observation, the location of the fixed point being block definition dependent, appropriate fine-tuning of any decimation free parameters can be crucial for achieving rapid approach in few steps.               

Here we employ two-lattice matching with RG block transformations implemented by the recursions (\ref{recur}) described above. They can be explicitly evaluated to any desired accuracy on lattices of any size, so no inherent limitations due to finite size arise. 
The blocked action resulting after each RG step is explicitly obtained, so it can used to 
ascertain approach to the RT and perform two-lattice matching. The transformations contain one parameter (cf. (\ref{decpar1}) - (\ref{decpar2})), which should be fixed 
for matching optimization. 

A basic feature of our decimations is that, regardless of the choice of the 
initial plaquette action, a single step suffices to generate an action 
of the form (\ref{actexp}) generally containing a large (infinite) set of representations. 
This is important as flow in such a large-dimensional interaction space makes it possible to avoid getting stuck at (finite-dimensional) lattice artifact boundaries.   
Furthermore, MCRG construction of blocked actions \cite{T1mcrg} shows that one-plaquette terms 
with a large number of characters are the most relevant action terms for long-scale dynamics. 
This is precisely the type of action resulting from our decimations, and may be the reason for their apparent efficacy in computing long-distance dynamics observables as seen below. 

\begin{figure}[ht]
\begin{center}
\includegraphics[width=0.6\textwidth]{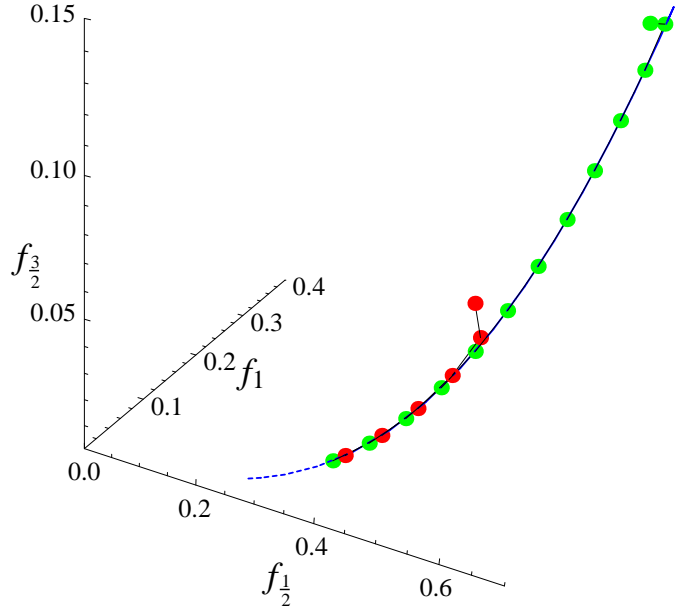}
\end{center}
\caption{Flow from the $SU(2)$ fund. Wilson action with $\beta=4$ (green dots) and $\beta=2.5$ (red dots) showing rapid approach to the RT. First three non-trivial (normalized) expansion  coefficients shown. \label{RGf1}}
\end{figure}

In the following the starting action for our decimations ($n=0$) will always be taken to be the fundamental representation Wilson action. (Other choices such as mixed actions containing 
several representations can be treated in exactly the same way.)
One finds that the flow under successive decimations reaches a unique RT irrespective of such a choice, though of course the number of steps needed to reach it depends on the initial point in action space. With the fundamental Wilson action as the starting action the approach is found to be very rapid as illustrated in Fig. \ref{RGf1}. 
%Flow from the $SU(2)$ fund. Wilson action with $\beta=4$ (green dots) and $\beta=2.5$ (red %dots). First three non-trivial expansion coefficients shown.  

The effective coupling (\ref{effcoupl}) provides a good way to label points along the RT. 
If, starting from some 
Wilson action coupling $\beta$, after $n$ steps the point $\beta^{(n)}(\beta)$ lies on the RT, subsequent RG steps generate a sequence of points $\beta^{(n+1)}$, $\beta^{(n+2)}, \cdots$ 
hopping along the RT.  With scale factor $b=2$, and for all large and intermediate values of $\beta^{(n)}$, the effective beta function is varying slowly enough 
for a linear interpolation to provide an excellent approximation to the RT points lying between pairs of neighboring points $\beta^{(n)}, \beta^{(n+1)}$. 

The decimations become exact in the $\beta\to \infty$ limit. Computation of the step scaling function (beta function) from the decimations in the weak coupling scaling region reproduces the perturbation theory prediction to within $2\%-3\%$.  We next use them to obtain 
critical couplings and string tensions by lattice matching for the $SU(2)$ and $SU(3)$ gauge theories.

\section{Critical couplings and string couplings by two-lattice matching} 
\subsection{Critical couplings} 

At physical temperature $T=1/aN_\tau$, lattice with temporal extent $N_\tau$, spacing $a$ and lattice with $N^\prime_\tau$, $a^\prime$ are related by:  
\beq
a^\prime = {N_\tau \over N^\prime_\tau } a \,.\qquad  \label{latspace2}
\eeq
If after blocking $n$ and $n^\prime$ times, respectively, the two flows 
reach the same point on the RT, using (\ref{latspace1}) this implies 
\beq
% a^\prime = b^{(n-n^\prime)} a \qquad     \Longrightarrow   \qquad  
   n - n^\prime = \log_b \left({N_\tau \over N^\prime_\tau}\right) \,. \label{latspace3} 
\eeq
So, at $T=T_c$ one has 
\beq 
\beta^{(n)}(\beta_c(N_\tau)) = \beta^{(n^\prime)} (\beta_c(N^\prime_\tau)) \,.\label{eqeffcr}
\eeq 
This suggests the following simple matching procedure. \\
 (i) Assuming  $\beta_c(N_\tau)$ known for one $N_\tau$, 
take 
\bea 
n& = & \log_b N_\tau  + m  \nonumber \\
n^\prime & = & \log_b N^\prime_\tau  + m \label{nchoice}  
\eea
Integer $m=0,1, \ldots$ is chosen so $n,n^\prime$ large enough to be on the RT. 
(If the so-chosen $n$ or/and $n^\prime$ turn out to be non-integer, one performs $[n]$ and $[n]+1$ steps, where $[n]$ is the nearest integer to the chosen $n$ from below, and uses interpolation 
for the RT points in-between as mentioned above.)\\
(ii) With $n$, $n^\prime$ and $\beta_c(N_\tau)$ given, solve (\ref{eqeffcr}) for $\beta_c(N^\prime_\tau)$. This means that the starting point of the flow on the $N^\prime_\tau$ lattice is adjusted 
to satisfy (\ref{eqeffcr}).  
\subsection{String tensions  ($T=0$)}
A similar procedure allows one to obtain string tensions by matching. 
Assume that two RG flows from starting Wilson action at $\beta_0$ and $\beta_1$ end up at the same RT point after $n_0$ and $n_1$ steps, respectively. Then one has 
\beq
\beta^{(n_0)}(\beta_0) = \beta^{(n_1)}(\beta_1)  \label{eqeffst} 
\eeq
and 
\beq
a_1\sqrt{\sigma} = b^{(n_0 - n_1)} a_0 \sqrt{\sigma}  \label{eqst}
\eeq 
Suppose $a_0 \sqrt{\sigma}$ known.  
Choose $n_0$ large enough to be on the RT. 
Determine $n_1$ so that (\ref{eqeffst}) is satisfied.
$ a_1\sqrt{\sigma}$ is then obtained directly from (\ref{eqst}).

\subsection{Numerical results}
It is important to maintain high accuracy in working with the expansions (\ref{exp}) under  
blocking iteration. For $SU(2)$ we typically use fifty group characters in the expansions (\ref{exp}). This implies for, say, $\beta=5$ omitted higher character coefficients $f_j=F_j/F_0$, and accompanying bounds on the series remainder, of the order of $10^{-45}$. For $SU(3)$ we truncate (\ref{exp}) at characters $j\equiv (p,q)$ with $p\geq 20$, $q\geq 20$; this implies remainders at $\beta=10$ of less than $10^{-12}$. 
%Errors due to truncation in the character expansions (\ref{exp}) are thus negligible. 

The scale factor is always taken to be $b=2$. The adjustable parameter in the decimation recursions (\ref{recur}) - (\ref{decpar2}) is $c$, which we set at $c=0.10$ in the case of $SU(2)$ and $c=0.24$ in the case of $SU(3)$. With no other parameters present, straightforward numerical evaluation of the recursion relations can then be carried out. 

We take one value of $\beta_c(N_\tau)$ from MC data, which serves to fix the 
scale and apply the procedure above to obtain critical coupling values 
for other lattices. Results for $SU(2)$ are shown in Table \ref{TablebetacrSU(2)}. Two sets of computed $\beta_c$ values are shown  in 
Table \ref{TablebetacrSU(2)} (columns 1 and 2) corresponding to two different choices of the MC data point (underlined entries). The table also shows comparison with the values obtained by MC simulation \cite{LTW} - \cite{Hetal}  in each case (column 3). The agreement is 
very good - at the $2\%-3\%$ level. 
Results for critical couplings in the $SU(3)$ gauge theory are displayed in Table \ref{TablebetacrSU(3)}.  Agreement with MC simulation data  \cite{LTW} 
% \cite{Langelage2.1,Lucini3.1}. 
is again very good, typically within a few percent.

\begin{table}[htb]
\centering
\begin{tabular}{|c|@{\hspace{0.5cm}}c@{\hspace{0.5cm}}|@{\hspace{0.5cm}}c@{\hspace{0.5cm}}|@{\hspace{0.5cm}}c@{\hspace{0.5cm}}|}
  \hline
  $N_\tau$ &  $\beta_c$ & $\beta_c$&$\beta_c$(MC) \\
  \hline
  3  & 2.1875 & 2.1957& 2.1768(30) \\
  4  & 2.2909 & \underline{2.2991}& 2.2991(02) \\
  5 & 2.3600 & 2.3683 & 2.3726(45)  \\
  6 & 2.4175 & 2.4258  & 2.4265(30) \\
  8 & 2.5097 & 2.5180 & 2.5104(02) \\
  12  & \underline{2.6355} & 2.6440 & 2.6355(10) \\
  16  & 2.7275 & 2.7361 & 2.7310(20)\\
  32  & 2.9487 & 2.9574   &   \\
  \hline
\end{tabular}
\caption{Critical couplings $\beta_c(N_\tau)$ for $SU(2)$ computed from lattice 
matching of decimations. Column 1 and 2 show the values obtained for two different choices 
(underlined entries) of the one data point taken from MC data (see text). Column 3 shows the values 
from MC simulations for comparison. \label{TablebetacrSU(2)}} 
\end{table}

\begin{table}[htb]
\centering
\begin{tabular}{|c|@{\hspace{0.5cm}}c@{\hspace{0.5cm}}|@{\hspace{0.5cm}}c@{\hspace{0.5cm}}|@{\hspace{0.5cm}}c@{\hspace{0.5cm}}|}
%{|c|l|l|l|}
 \hline
 $N_\tau$ & $\beta_c$ & $\beta_c$ & $\beta_c$(MC) \\
 \hline
  4  & 5.6501 & 5.6329 & 5.6925(002) \\
  6 & \underline{5.8941} & 5.8773  & 5.8941(005) \\
  8 & 6.0773 & 6.0595 & 6.0010(250),6.0625(18)  \\
  10  & 6.2018 & 6.1837 & 6.1600(70)\\
  12 & 6.3084 & 6.2900  & 6.2680(120),6.3385(55) \\
  14  & 6.4015 & \underline{6.3830} & 6.3830(100)\\
  16  & 6.4845 & 6.4658 & 6.4500(500)\\
  32  & 6.9024 & 6.8829 &  \\
 \hline
\end{tabular}
\caption{Critical couplings $\beta_c(N_\tau)$ for $SU(3)$ 
computed from lattice matching of decimations and comparison with MC simulation data.
Same format as in Table 1.}
\label{TablebetacrSU(3)}
\end{table}
String tensions in $SU(3)$ obtained by the method above are displayed in the same format in Table \ref{TableSTSU(3)}. Very similar results are obtained for $SU(2)$ \cite{CT}. 
Good agreement with MC data \cite{Hetal} - \cite{EHK} 
is again obtained in all cases. 

Consideration of fermionic observables by similar RG recursion methods is a rather more demanding 
proposition. Some preliminary attempts are reported in \cite{CT1}. 

\begin{table}[ht]
\centering
\begin{tabular}{|c|@{\hspace{0.5cm}}c@{\hspace{0.5cm}}|@{\hspace{0.5cm}}c@{\hspace{0.5cm}}|@{\hspace{0.5cm}}c@{\hspace{0.5cm}}|}
%{|l|l|l|l|}
  \hline
  $\beta$ & $a\sqrt \sigma $ & $a\sqrt \sigma $  & $a\sqrt \sigma $(MC) \\
  \hline
  5.54 & 0.5580 & 0.5878  & 0.5727(52) \\
  5.6  & 0.5070 & \underline{0.5295}  & 0.5295(09), 0.5064(28)\\
  5.7  & 0.4205 & 0.4264  & 0.4099(12), 0.3879(39)\\
  5.8 & 0.3486 & 0.3508   & 0.3302(15) \\
  5.9  & 0.2919 & 0.2931 & 0.2702(19) \\
  6.0   & 0.2465 & 0.2433  & 0.2269(62), 0.2209(23)\\
  6.2  & 0.1698 & 0.1671  & 0.1619(19), 0.1604(11)\\
  6.4 & \underline{0.1214} & 0.1180  & 0.1214(12), 0.1218(28) \\
  6.5 & 0.1010 & 0.0983  & 0.1068(09) \\
  6.8 & 0.0616 & 0.0599  & 0.0738(20) \\
  \hline
\end{tabular}
\caption{String tensions $a\sqrt \sigma $ for $SU(3)$ computed from lattice matching of decimations. Same format as in Table 1. \label{TableSTSU(3)}}
\end{table}
%\section{Conclusions}
\vspace{0.3cm}
This work was partially supported by the NSF under NSF-PHY-0852438.

\end{document}